\documentclass[10pt, conference, letterpaper]{IEEEtran}

\IEEEoverridecommandlockouts
\usepackage{cite}
\usepackage{amsmath,amssymb,amsfonts}

\usepackage{mathptmx}
\usepackage{textcomp}
\usepackage{xcolor}
\usepackage{url} 

\usepackage{caption}
\usepackage{graphicx, subfigure}
\graphicspath{{./graphics}}
\usepackage[ruled,linesnumbered]{algorithm2e}
\SetKwIF{If}{ElseIf}{Else}{if}{}{else if}{else}{}
\SetKwFor{While}{while}{}{}
\SetKwFor{For}{for}{}{}
\RequirePackage{CJKnumb}
\usepackage{multirow} %multirow for format of table 
\usepackage{amsmath} 
\usepackage{xcolor}
\usepackage{verbatim}
\usepackage{mathtools}
\usepackage{xspace}
\usepackage{float}
\usepackage{subcaption}
\newcommand{\yolo}{\texttt{YOLOV3}\xspace}
\newcommand{\sys}{\texttt{TargetFuse}\xspace}

\newcommand{\mwx}[1]{\textbf{\color{red}{#1}}}

\def\BibTeX{{\rm B\kern-.05em{\sc i\kern-.025em b}\kern-.08em
    T\kern-.1667em\lower.7ex\hbox{E}\kern-.125emX}}
\begin{document}

\title{Resource-efficient In-orbit Detection of \\Earth Objects
%{\footnotesize \textsuperscript{*}Note: Sub-titles are not captured in Xplore and
%should not be used}
%\thanks{Identify applicable funding agency here. If none, delete this.}
}
\author{\IEEEauthorblockN{Qiyang Zhang$^{1}$, Xin Yuan$^{1}$, Ruolin Xing$^{1}$, Yiran Zhang$^{1}$, \\  Zimu Zheng$^{2}$, Xiao Ma$^{1*}$, Mengwei Xu$^{1}$, Schahram Dustdar$^{3}$, Shangguang Wang$^{1}$\\ \small
{$^{1}$State Key Laboratory of Networking and Switching Technology,
BUPT}, Beijing, China\\
{$^{2}$Huawei Technologies Co., Ltd} \quad\\ {$^{3}$Distributed System Group, TU Wien}, Vienna, Austria\\
\\
\{\tt \small {qyzhang;bupt\_yx,xrl,yiranzhang,maxiao18,mwx,sgwang\}@bupt.edu.cn};\\dustdar@dsg.tuwien.ac.at;
zimu.zheng@huawei.com}

}

%\author{\IEEEauthorblockN{Anonymous Author}
%\IEEEauthorblockA{\textit{dept. name of organization (of Aff.)} \\
%\textit{name of organization (of Aff.)}\\
%City, Country \\
%email address or ORCID}
%}
%\{\tt \small {qyzhang;bupt\_yx,xrl,yiranzhang,maxiao18,mwx,sgwang\}@bupt.edu.cn};\\dustdar@dsg.tuwien.ac.at;
%zimu.zheng@huawei.com}
\maketitle
\renewcommand{\thefootnote}{\fnsymbol{footnote}}
\footnotetext[1]{Corresponding author.\\
This work is supported by NSFC (No.62032003, 61921003, 62372061, 62302055, 62102045), Beijing Nova Program (No.Z211100002121118), and Young Elite Scientists Sponsorship Program by CAST (No.2021QNRC001).}

%%
%% The abstract is a short summary of the work to be presented in the
%% article.
\begin{abstract}
%The decreasing costs of deploying satellites and advancements in satellite technology have led to an emergence of large satellite constellations. 
With the rapid proliferation of large Low Earth Orbit (LEO) satellite constellations, a huge amount of in-orbit data is generated and needs to be transmitted to the ground for processing.
However, traditional LEO satellite constellations, which downlink raw data to the ground, are significantly restricted in transmission capability.
%Traditional satellite systems operate under the bent-pipe architecture, where satellites downlink raw observations to ground stations. However, this becomes challenging due to the large quantity of images and short-term ground station contacts. 
%Orbital edge computing (OEC) is a novel solution that can relieve the downlink issue, but it also shifts the challenge to the inelastic computational capabilities onboard satellites with lower energy. 
Orbital edge computing (OEC), which exploits the computation capacities of LEO satellites and processes the raw data in orbit, is envisioned as a promising solution to relieve the downlink burden.
Yet, with OEC, the bottleneck is shifted to the inelastic computation capacities.
The computational bottleneck arises from two primary challenges that existing satellite systems have not adequately addressed: the inability to process all captured images and the limited energy supply available for satellite operations.
In this work, we seek to fully exploit the scarce satellite computation and communication resources to achieve satellite-ground collaboration and present a satellite-ground collaborative system named \sys for onboard object detection.
%but it also shifts the challenge to the inelastic computational capabilities onboard satellites with lower energy.
%In this work, we present \sys, a satellite-ground collaborative system that addresses both computational and downlink bottlenecks across the satellite and ground. 
%这里少一句总结说明你提的这个架构main idea是什么 如何同时克服这两个bottleneck的 不能上来就介绍这个架构的细节
\sys incorporates a combination of techniques to minimize detection errors under energy and bandwidth constraints.
Extensive experiments show that \sys can reduce detection errors by 3.4$\times$ on average, compared to onboard computing. \sys achieves a 9.6$\times$ improvement in bandwidth efficiency compared to the vanilla baseline under the limited bandwidth budget constraint.

\end{abstract}
%\title{Some Title{\vspace-2em}}

\section{Introduction}
Earth-observation (EO) satellites collect multispectral images for geospatial analysis, providing valuable sensing and computational applications in the harsh space environment, characterized by highly constrained energy and network connectivity.
Visual tasks, which detect vehicles along interstate highways to estimate traffic \cite{liu2016highway}, count buildings from key areas to predict population \cite{han2022context}, or monitor animals from the wildness to track their behaviors \cite{hodgson2016precision}, etc, are among the key use cases for EO satellites. 
%Communication and computation ability constrain satellite utility.
Advances in technology enable satellites to collect vast amounts of Earth images daily, often reaching tens of Terabytes \cite{nasa2015esds, doug2020teraspace}. However, traditional satellites operate as “bent-pipe” and typically downlink all raw observations to the ground, which is significantly restricted in transmission capability due to the scarce bandwidth resources (e.g., tens of Mbps) and limited satellite-ground connection duration.

%As an effort to support analytics in the satellite scenarios, this work focuses on in-orbit detection of Earth objects.

%这句话出现得太早了 这一段还只是要介绍背景 建议删掉 第一段直接进入正题，介绍完EO satellite之后，给出一些数据，说明visual tasks会产生非常大量的数据量，后面介绍传统EO卫星的visual tasks通过弯管架构 数据传输有XXX问题；第一段的落点就落在弯管架构无法把数据都下载到地面处理

\begin{comment}

Counting objects is one of the significant EO use cases, including counting vehicles along interstate highways to estimate traffic \cite{liu2016highway}; counting buildings from key areas to predict population \cite{han2022context}; counting animals from the wildness to track their behaviors and therefore monitor their distribution \cite{hodgson2016precision}, etc.
As an effort to support analytics in the satellite scenarios, this work focuses on in-orbit counting of Earth objects. 
Figure \ref{fig:1234} shows that satellites utilize neural network (NN) counters to count objects in orbit. The object counts are aggregated for captured images on each track, and satellites emit the aggregated counts with higher confidence. When satellites come into contact with ground stations, they downlink parts of the images to the ground for further analysis.
\end{comment}
\begin{figure}[t]
   \centering
   \setlength{\belowcaptionskip}{-0.5cm}
   \begin{center}
     \includegraphics*[width=\linewidth]{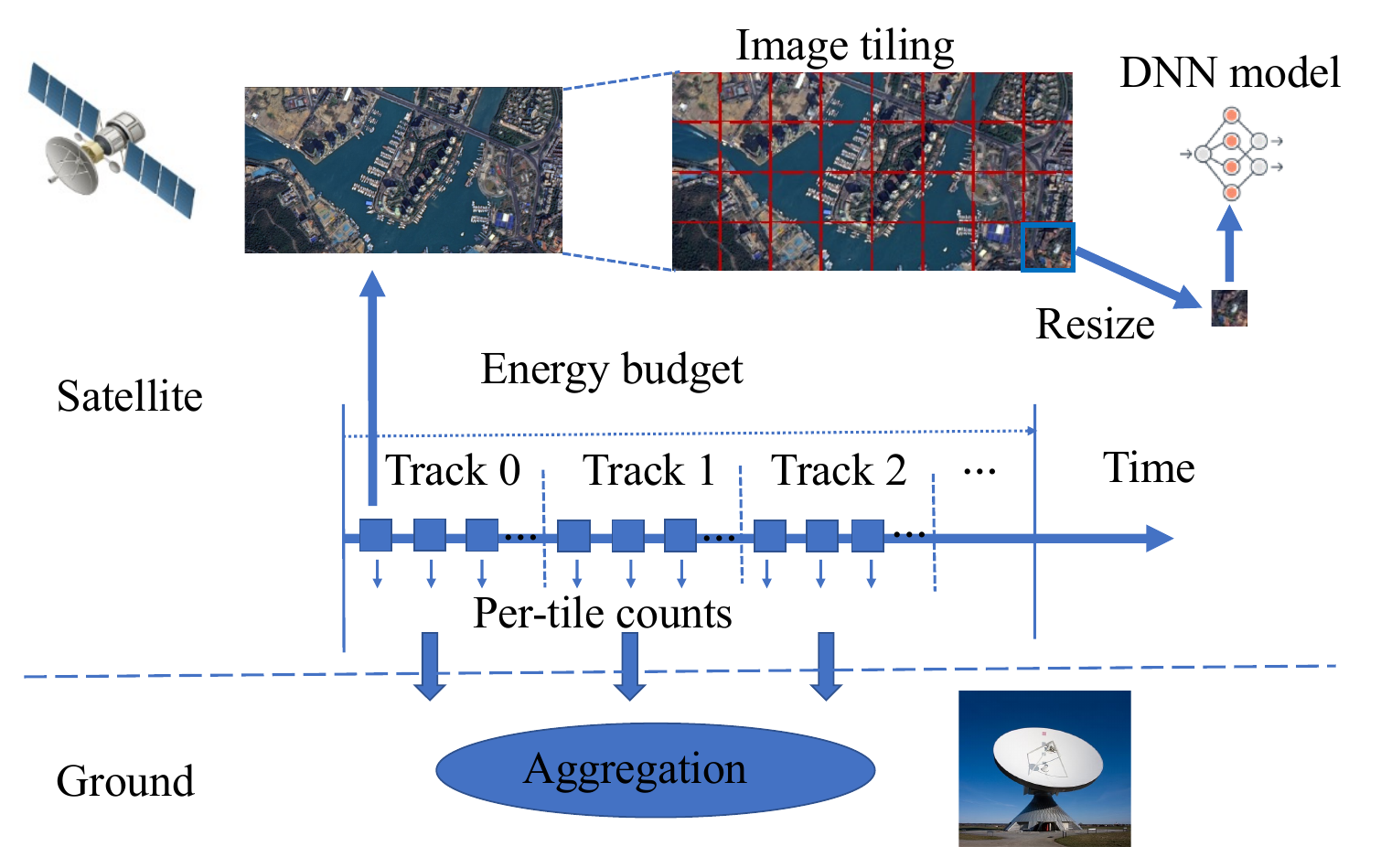}
     %\vspace{-1.5cm}
   \caption{A satellite periodically captures Earth images. Images are pre-processed into tiles before applying DNN models.}
   \label{fig:1234}
   \end{center}
   %\vspace{-6cm}
\end{figure}

%\new{recheck}
%合并到第一段重新组织一下

%第二段以为了解决bent-pipe的问题开始
To overcome the limitations of the bent-pipe architecture, Orbital Edge Computing (OEC) has emerged as a promising solution, which mitigates the communication bottleneck by processing the observation data in space and sending the process results to the ground rather than directly downlinking the raw data\cite{denby2020}. 
However, in-orbit processing faces a computational bottleneck due to the inherent limitations of satellites, including both computational capacity and constrained energy. 
%感觉这里要表达的不是challenging吧？感觉和上一段结尾是一个意思 就是更详细地解释了 这一段主要是具体解释in-orbit processing需要的算力很大，超出了卫星计算能力的限制 这里可以给出一些数据？说明计算容量和energy是如何受限的 和 
%Processing in-orbit raw observations is also challenging. %这句话建议去掉，下面的合并到上一段，重新组织
EO satellite images are large (i.e., hundreds of millions of pixels) and arrive at a high rate, far exceeding what embedded satellite hardware can process.
The computational bottleneck arises not only from lower computing power hardware but also from the inability to process all images within the energy budget harvested from the less productive solar panel.
For instance, Baoyun satellite can harvest up to 260KJ of energy daily, but not all of it is utilized for computing. When about 150KJ of energy is allocated for computing, satellite can only support the computation of about 22\% of the observable high-resolution 3K satellite images.
%As shown in Figure \ref{fig:1234}, satellites run Deep Neural Network (DNN) models in orbit to detect objects. 
%The detection results are aggregated for the captured images on each track, and satellites emit the aggregated count results.
Therefore, as an effort to support analysis for EO, this work focuses on in-orbit detection of Earth objects, considering the inherent limitation in computational and energy.
Fig.~\ref{fig:1234} shows that satellites utilize Deep Neural Network (DNN) models in orbit to detect objects. Detection results for images captured on each track are aggregated, and satellites transmit these aggregated counts.%When satellites come into contact with ground stations, they downlink parts of the images to the ground for further analysis.}

%下面这段怎么介绍exsiting work和一些说明揉到一起了？ 把一些数据的说明移到前面 这一段纯粹说明existing work有什么问题（地面对图像处理的方案直接放星上会怎么样？星上现有的方案有什么问题，然后引出你的架构）
Terrestrial applications typically divide large images into smaller images and process each image on DNN models.
However, they focus only on improving application performance, ignoring the vital system factor of computational overhead \cite{van2018you}.
Prior attempts to address the computational bottleneck focused on distributing in-orbit processing across a constellation aimed at a particular purpose \cite{denby2020}, but the relatively high-cost solutions did not considered the energy budget of each satellite.
In-orbit computing still cannot meet onboard application requirements due to energy constraints. 
Therefore, a collaborative satellite-ground system is needed, where embedded satellite hardware initially processes images under energy budget constraints and transmits crucial images to the ground despite facing limited bandwidth \cite{desai2020power}. %不对啊 星地协同系统 卫星图像是要下传到地面的？这不是弯管架构
In situations where a downlink system encounters a scarce link, efficient transmission within a short-term contact time should be prioritized. However, today's satellites transmit data indiscriminately \cite{wang2023tiansuan}. 
Less attention has been paid to the scarce downlink bandwidth budgets.
Therefore, this work centers on processing images onboard the satellite and selectively transmitting crucial images by utilizing the downlink capability.%communication的不是已经介绍完了吗 怎么又重复了一遍 没太明白这句话的作用

%前面介绍星地协同系统的shortage不是很明确，感觉需要加几句话
%前面没说challenging啊 这不叫challenge 这是shortage 
There are potential opportunities to deal with these computational and downlink bottlenecks in the collaborative satellite-ground system:
%后面三条的opportunity先用简短的一句话总结这个opportunity具体是什么，后面再解释
%(1) Satellite images present an opportunity for division into tiles and each tile is scaled to match the input size of in-orbit DNN models. Our extensive experiments on widely-used datasets revealed the impact of tile size on both detection accuracy and computing overhead (execution time). By balancing the two factors, we determine the optimal tile size for the DNN model execution;
(1) By tiling large images into smaller ones and resizing them to fit standard DNN models' input size, there exists an optimal tile size that satisfies higher detection accuracy and lower computational overhead.
Preliminary experiments conducted on widely-used datasets reveal that tile size affects both detection accuracy and computing overhead (execution time). Therefore, we propose finding the optimal tile size, balancing accuracy and computational overhead, for executing the onboard DNN model;
%第一个opportunity具体是什么？
(2) Optimal tile size also establishes the confidence thresholds for onboard DNN models, influencing the decision of whether to downlink corresponding tiles to the ground. To efficiently utilize the available bandwidth, we implement bandwidth-aware downlinking throttling to dynamically select and downlink crucial tiles within the given bandwidth budget. The selection process will be based on the confidence thresholds set by the onboard DNN models;
%thereby ing the computation and communication ; %这opportunity？ 
(3) Satellite images contain some semantically geospatial feature contexts with a high degree of similarity, specifically for the captured images as the satellite passes over its ground track. To alleviate computational and downlink bottlenecks, 
we introduce a lightweight, clustering-based data deduplication technique that leverages geospatial feature contexts. This technique optimizes data processing by efficiently identifying and removing redundant data, thereby reducing computational and downlink burdens.

%介绍opportunity和你基于这个opportunity的设计分开说 前面介绍完opportunity之后 后面直接说明你提出了一个XXX架构 这个架构具体idea是什么，是如何借助这些opportunity进行设计的

We present \sys, a collaborative satellite-ground system under the inherent limited computational and downlink constraints. %designed to produce aggregated object counting, 前面也在说星地协同系统 这里也叫星地协同系统 看不出特色来
 %是bandwidth吗？不是computation吗
To ensure a realistic simulation, we collect new data, including satellite operation details such as the computing power of embedded hardware, based on tested in-orbit satellites.
We provide a comprehensive evaluation of \sys across various energy budgets, embedded hardware setups, and bandwidth budgets.
Extensive experiments show that \sys can reduce detection errors by 3.4$\times$ on average, compared to onboard computing. \sys showcases a remarkable 9.6$\times$ improvement in bandwidth efficiency compared to the vanilla baseline under the limited bandwidth budget constraint.
%\sys employs a combination of novel techniques, including selecting an optimal tile size by balancing detection accuracy and computing overhead, generating geospatial contexts for the deduplication of similar images during processing or downlinking, and 
%selecting tiles with appropriate confidence thresholds for downlinking under a saturated downlink.%感觉直接说你的架构是几个技术的combination把你的工作削弱了，你可以分开介绍，为了克服XXX，你采用了XXX技术，达到了什么效果
%前面要提一下你的方案是怎么验证的 结果如何
Our contributions can be summarized as follows:%你的第一条contribution怎么感觉前面都没出现过？contribution总结这里的内容在introduction部分都应该出现过才行
\begin{itemize}
\item We design and implement a satellite-ground collaboration system for object detection, a critical use case for EO satellite, aiming to minimize detection errors. To ensure the system's realism, we collect and publish new data that contains satellite operation details\footnote{\url{https://www.tiansuan.org.cn/}}.
%Once published, we will open source our system and data.
%这段第一句话句式不对啊 不能是XXX是XXX，应该是你提出了XXX ，你构建了XXX 或者你研究了XXX这种才是contribution,收集到的数据是你的contribution吗
%所以你是构建了一个星地协同系统？那你这段应该是We implement/build/design a satellite-ground collaboration system

\item We propose image tiling to balance detection accuracy and computational overhead, clustering-based data deduplication to alleviate the computational bottleneck, and bandwidth-aware down-linking throttling to address downlink bottlenecks.%太笼统了 显得很平淡，再深入具体一点

\item We conduct extensive experiments across various energy budgets, embedded hardware configurations, and bandwidth budgets, and demonstrate that \sys's superior performance against four baselines.

\end{itemize}

%这是什么？这种放到第一条contribution里 你就说你设计了一个架构 这个架构是第一个
%To our knowledge, \sys is the first software satellite-ground system under the limited energy budget constraint. 
%Our work is dedicated to advancing the geographic frontier of satellite image analysis.%这种话太中式了 删掉

\section{Background and Motivation}
\begin{comment}
\mwx{
Look at that ASPLOS paper.
First, describe the traditional way of remote sensing: sending everything to the ground (bent pipe).
Then, demonstrate this approach is not efficient due to limited satellite-ground bandwidth (show numbers).
Next, describe the trends of COTS hardware in space that enables in-orbit processing (as an opportunity and motivation for this work).
Finally, describe the in-orbit constraint: energy budget (challenge of this work). In-orbit alone is not enough, so we need satellite-ground collaboration.
}
\end{comment}
\subsection{EO Satellites}
EO satellites collect raw sensor data for geospatial analytics.
These satellites capture images along their ground track, generating geospatial images that cover hundreds of square kilometers and contain hundreds of millions of pixels. 
The level of detail present in these images is described by the ground sample distance (GSD) \cite{denby2020}, which is determined by orbit altitude, sensor size, and camera characteristics \cite{denby2020}. 
Besides, the high velocity of satellites, up to 7.9 km/s, results in brief periods of visible contact with ground stations. These periods typically last only a few minutes, sometimes less than 8 minutes, and may occur infrequently.
\begin{comment}
Currently, in bent-pipe architectures \cite{denby2020orbital}, the majority of Earth-observation satellites transmit raw observations to ground-based ML algorithms for processing. 
\end{comment}

During a single orbit revolution, a single satellite captures more images than it is capable of downlinking \cite{denby2023kodan}. This is due to the downlink capacity of current satellite sensors, which is insufficient to support their data rates. 
 Currently, the majority of Earth-observation satellites are organized in a bent-pipe architecture \cite{denby2020}, where raw observations are transmitted to ground stations and then processed by machine learning algorithms.
However, satellites only achieve downlinking rates of no more than hundreds of Mbps using Ka-band \cite{nasa2022sotasat}, resulting in a limited daily downlink data volume. For instance, given that a contact session lasts for 6 minutes, the system can downlink a maximum data quantity of 4.39 GB at a downlinking speed of 100 Mbps \cite{oo}. The downlink bottleneck prevents these daily global observations from being transmitted to the ground.
%Bent-pipe satellites waste limited downlink bandwidth by indiscriminately transmitting raw observations containing an amount of low-value data.
%For instance, 67\% of the observable data obscured by clouds and low-value to most customers is downlinked with a bent-pipe \yiran{cite?}. 
What's more, not all raw observations contain high-value data. Bent-pipe satellites may waste the precious downlink bandwidth as they indiscriminately transmit raw observations independent of the value of data. Statistics show that 67\% of the observations are obscured by clouds, thus becoming low-value to users \cite{li2021towards}.

%\yiran{The overall logic of 2.1 is not very clear. }

%\yiran{1st paragraph: earth-observation; 2nd paragraph: periods and low value data; 3rd paragraph: downlink capacity is insufficient}

%\yiran{Suggestion:->}

%\yiran{1st paragraph: earth-observation and periods; 2nd paragraph: downlink capacity is insufficient (thus is limited and precious); 3rd paragraph: what's more, not all observations are high-value data (thus low-value data may waste precious bandwidth); }

\subsection{Orbital Edge Computing}
\begin{comment}
    
Recently, OEC \cite{denby2019orbital,denby2020orbital} has been developed to address the downlink bottleneck, in which satellites process raw data in space. OEC addresses the limitations of the bent-pipe architectures \cite{denby2020orbital} distributing processing across a constellation.  Each commercial satellite constellation today is equipped with hundreds of high-datarate cameras, sensors ,and COTS hardware \cite{leung2018adcs}.
The satellite is lightweight, small-sized, and expensive, with each weighing a few kilograms, measuring a few centimeters, and costing millions of USD. However, the high cost does not provide highly-capable onboard processors. Currently available space-grade processors are often decades-old, “flight heritage”.
The satellite system must operate for decades in the space environment, which means that COTS hardware must be low-risk and highly reliable at the expense of performance.
Recent trends in space system have started to consider COTS embedded systems \cite{lovelly2017comparative}, which enables use of
in-orbit processing. Applying these terrestrial techniques directly to space is appealing, but computational capacity is subject to the
unique operating constraints. Unlike to Earth, all energy in space for expenditure must be harvested from the solar panel source, which is backed by a rechargeable energy buffer.
\end{comment}

OEC \cite{denby2020} has been proposed to address the downlink bottleneck, in which satellites process raw data in space. OEC aims to address the limitations of “bent-pipe” architectures \cite{denby2020} by distributing processing across a constellation. Nowadays, each satellite can be equipped with hundreds of high-datarate cameras, sensors, and commercial, off-the-shelf (COTS) hardware \cite{leung2018adcs}. The satellite is lightweight, small-sized, and expensive, with each satellite weighing a few kilograms, measuring a few centimeters, and costing millions of USD.

However, satellites still face limitations in providing highly-capable onboard processors. Currently, available space-grade processors are often decades-old, “flight heritage”. The satellite systems may operate for decades in the space environment, which means that COTS hardware may be low-risk and highly reliable at the expense of performance.
Hence, recent trends in space systems have started to consider COTS-embedded systems \cite{lovelly2017comparative}, which enable the use of in-orbit processing. Applying these terrestrial techniques directly to space is appealing, but computational capacity is subject to unique operating constraints. For instance, unlike on Earth, all energy expended in space must be harvested from solar panels, which is backed by a rechargeable energy buffer \cite{yang2022energy}. In line with typical 3U Cubesat systems \cite{denby2020}, the size of the tested satellite limits the area of the solar panel to 57.2 × 20.6 cm and thus limits the power to a range of 34-118 W.

\begin{figure}[t]
\centering
\subfigure[Physical map of Baoyun satellite \cite{baoyun,wang2023tiansuan}.]{
\begin{minipage}[t]{0.45\linewidth}
\centering
\includegraphics[width=\linewidth]{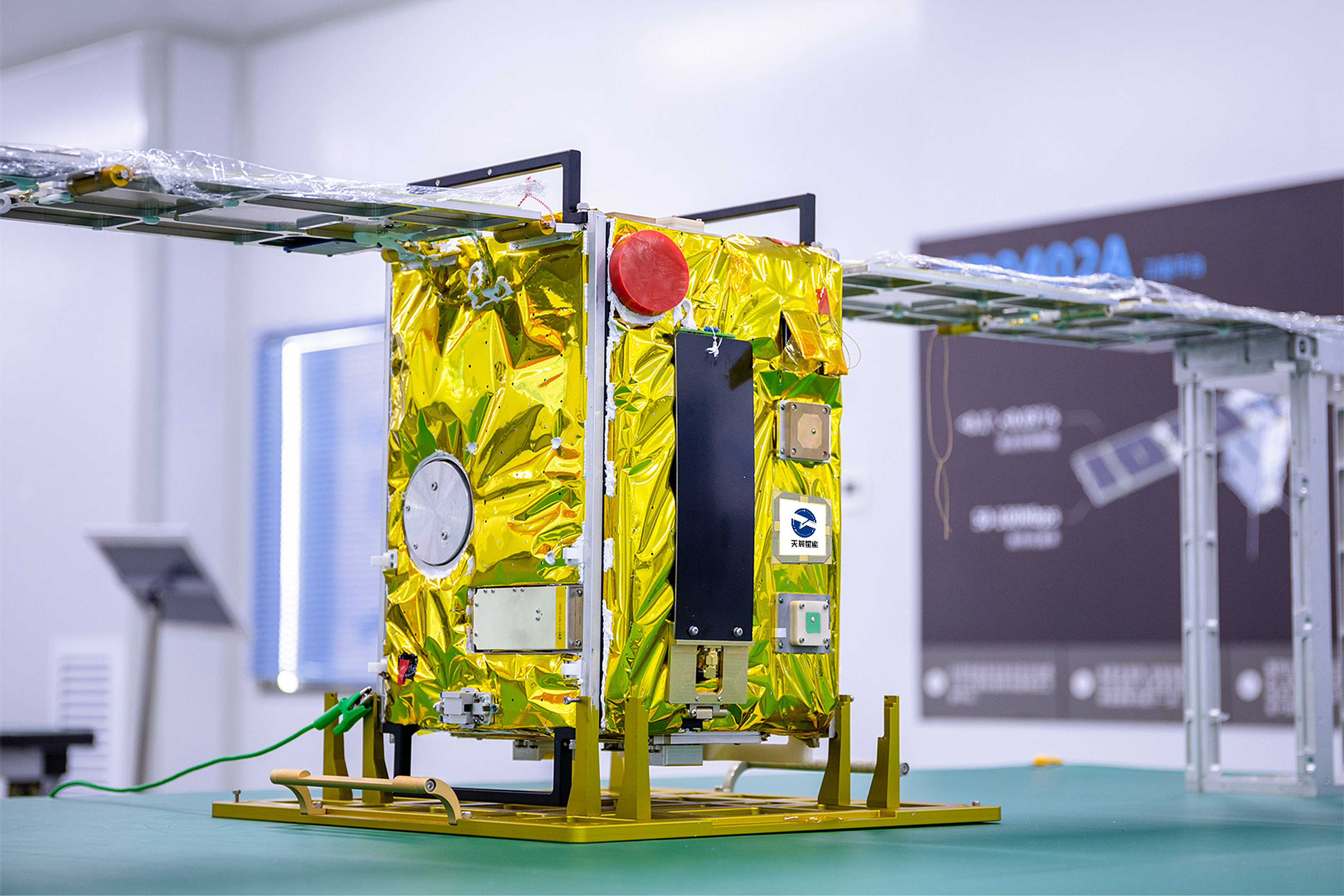}
%\caption{xView}
\end{minipage}%
}%
\subfigure[The distribution of energy expenditure in satellite operation.]{
\begin{minipage}[t]{0.45\linewidth}
\centering
\includegraphics[width=\linewidth]{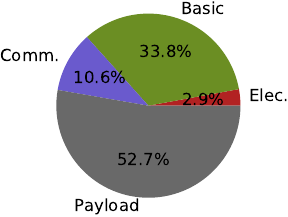}
%\caption{UAVOD10}
\end{minipage}%
}%
\centering
\caption{The real-world data, including energy and computing power, is collected from Baoyun satellite.}
\label{fig:90}
\end{figure}

The limited and inelastic computational resources also pose a major space systems challenge.
%\yiran{Question: "the satellite"  is the earth-observation satellite?}
Though the satellites are exposed to sunlight for about 60\% of each orbit period (e.g., approximately 90 minutes) \cite{davoli2019small}. Only 30\% of the solar energy is initially converted into battery power, and less than 50\% of the battery capacity is utilized for daily satellite operation over its lifetime \cite{li2021towards}.
We collected real-world data from Baoyun satellite and observed that computing in operation accounts for approximately 50\%, compared to other subsystems such as basic and electrical operations, as depicted in Fig.~\ref{fig:90}.
Therefore, during runtime, operating system (OS) of the satellite allocates an energy budget to the computing modules by adjustmenting in input performance and energy parameters. 
%This approach allows our system \yiran{this is in background. what is our system?} to run multiple applications in space, rather than being limited to a single application or being the only one running.

\subsection{Challenges for EO Computing} 
EO computing is a crucial application of OEC. However, two challenges arise at the orbital edge: a downlink bottleneck that hinders the transmission of all raw data and a computational bottleneck that restricts the processing of all data in orbit.
OEC has been proposed to address computational needs by distributing tasks across a constellation.
%Although effective in reducing per-satellite compute workload to meet full ground track coverage, this approach requires a large pipeline population. This approach incurs high monetary costs and is typically designed for vertically-integrated constellations deployed for a single purpose. 
Although effective in reducing per-satellite compute workload to meet full ground track coverage, this approach is in nature designed for vertically-integrated constellations for a single purpose, which requires a large pipeline population and incurs high monetary costs.
%Existing OEC work provides no technique to reduce the per satellite computational workload without increasing the constellation population. 
Existing OEC work can hardly reduce the per-satellite workload without increasing the constellation population \cite{denby2020}; this shortcoming is a key motivation for our work.

Processing raw observations in space presents significant challenges due to limited computational resources. Satellites have an energy limit that prevents them from processing all images, which creates a computational bottleneck that limits OEC's ability to address the downlink bottleneck. Unfortunately, 
improving the computational capacity of COTS hardware is difficult because of physical constraints. 
And there are no feasible options for adjusting the computational capacity of the hardware already in space. Furthermore, in-orbit computing alone is insufficient because each satellite has natural constraints such as volume, mass, and energy that prevent it from processing all images.

Compared to the space environment, a more collaborative approach between satellite and ground station is feasible by utilizing the relatively more favorable computing capacity and higher energy availability of ground stations. As a result, we consider the two bottlenecks of in-orbit processing jointly to adapt the computing hardware of the target satellite in space.

%\yiran{The overall logic of 2.3 is not very clear.}

%\yiran{The logic between the 2nd paragraph and the 3rd paragraph: what is the relationship between increasing constellation population and the two challenges?}

%\yiran{Suddenly mention earth-observation counting? Should first mention that  earth-observation counting is a vital/important orbital edge computing/earth-observation application}
\section{System Design}

\begin{figure}[t]
   \centering
   \setlength{\belowcaptionskip}{-0.5cm}
   \begin{center}
     \includegraphics*[width=\linewidth]{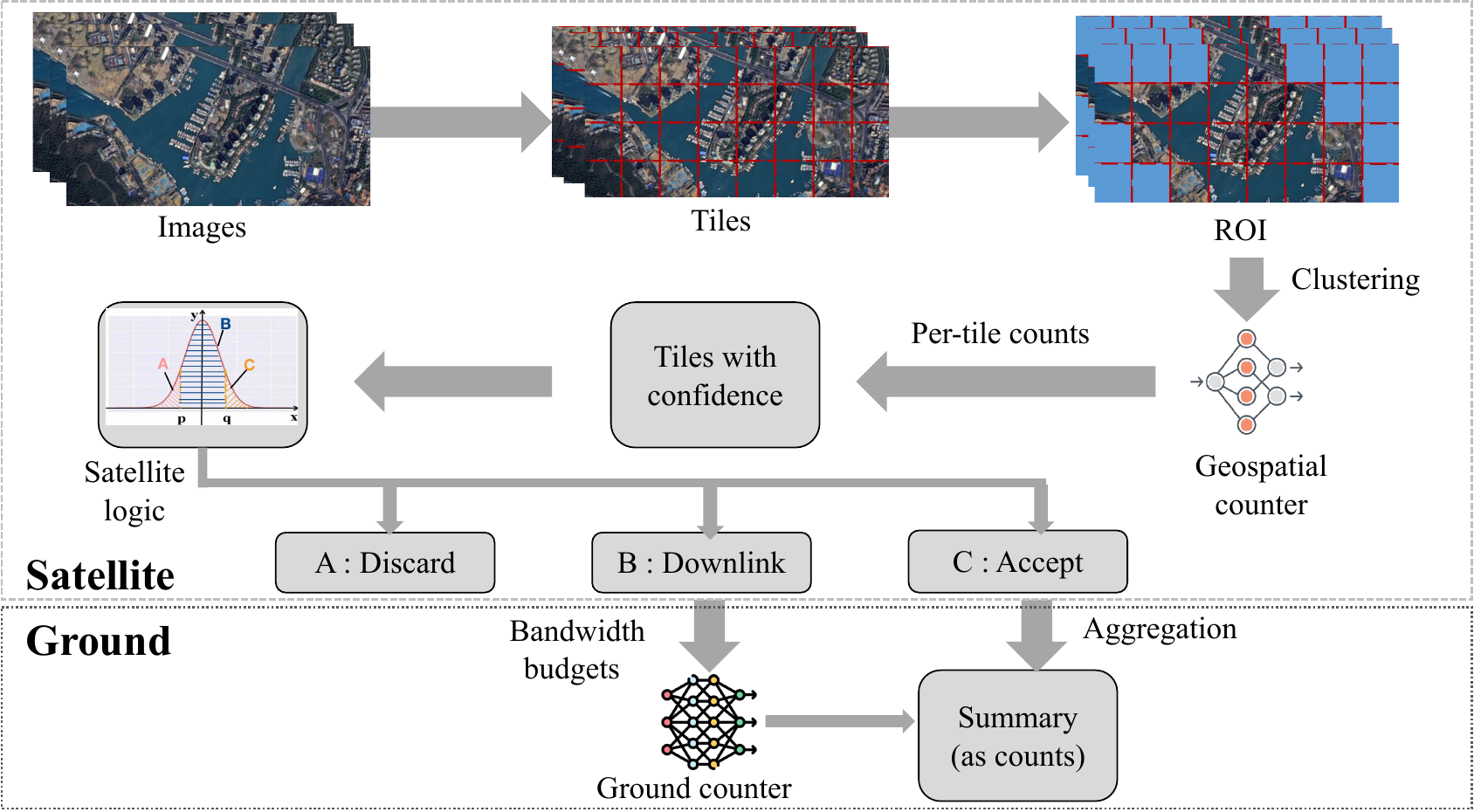}
     %\vspace{-1.5cm}
   \caption{The overall workflow of \sys.}
   \label{fig:123}
   \end{center}
\end{figure}

This work introduces a satellite-ground collaborative system tailored to counting, aiming to minimize counting errors in the challenging satellite environment. 
\sys executes a geospatial counter (e.g., a shallower DNN) in space with lower computing power, producing less accurate counts; and a ground counter (e.g., a deeper DNN) on the ground with higher computing power, producing more accurate counts.
To meet the computational and downlinking needs of counting applications, \sys leverages three techniques: adaptive image tiling, clustering-based data deduplication, and bandwidth-aware downlinking throttling. The workflow of \sys is shown in Fig.~\ref{fig:123}.
For each satellite image, \sys divides it into several tiles based on the image resolution and input size of the DNN counters.
Next, \sys automatically performs clustering-based data deduplication considering the similarity of the tiles. After that, \sys 
applies selection logic to determine which logic according to confidence thresholds from the onboard DNN counter. Note that onboard system also downlinks the counting result in space with vital tiles. Hence \sys emits the aggregated object count across space and ground.

\subsection{System Operation}
\subsubsection{Energy expenditure}~\sys performs orbital counting while adhering to the allocated energy budget of satellites along their trajectories. Utilizing data obtained from an in-orbit satellite \cite{wang2023tiansuan}, energy allocation extends beyond fundamental satellite operations, including propulsion and avionics. Energy is allocated for the following computing activities associated with counting: (1) $E_{cap}$ for capturing images; (2) $E_{com}$ for executing counting on images; (3) $E_{agg}$ for deriving aggregated counting results in space; (4) $E_{down}$ for downlinking the satellite images to be counted on the ground. The most energy-intensive activities (2) and (4) account for over 60\% of the total energy consumption, as illustrated in Fig.~\ref{fig:90}: during each orbital track, satellites perform several trillions of FLOPs and downlink some images to the ground. In contrast, activities (1) and (3) consume negligible energy: (1) only involves capturing thousands of images from the onboard camera, and (3) only involves a few hundred arithmetic operations. Therefore, activities (2) and (4) are the focus of this work \cite{xing}. These two activities align with the satellites lifetime design \cite{wang2023tiansuan,chen2023energy}, which optimally utilizes less than 50\% of the available battery energy. This efficiency allows for the estimation of the daily energy budget.

\subsubsection{Selection logic with different confidence thresholds.}~Prior to system execution, the satellite's OS captures numerous images and selects a DNN counter. The deployment of DNN counters on satellites has become increasingly crucial for guaranteeing counting accuracy. When selecting a counter, a confidence threshold is established based on the onboard satellite's DNN counter detection. This confidence threshold indicates the probability of accurately counting the objects and falls within the range [0,1] \cite{wenkel2021confidence}.

\subsubsection{Objective: Minimizing overall counting error while optimizing energy and bandwidth expenditure.}~\sys's objective is to minimize overall counting errors by allocating energy and bandwidth efficiently. In our implementation, the overall counting error is defined as the mean of the counts across all tiles, a widely employed metric in diverse applications \cite{agrawal2017low}. A smaller counting error indicates heightened confidence in counting accuracy, which ultimately translates to greater benefits for customers.
To address the computational and downlink bottlenecks in the counting application, \sys leverages the three following techniques.

\subsection{Adaptive Image Tiling}
For each image, \sys is designed to divide images into tiles with a comparatively lower execution overhead.
Processing large satellite images, typically containing thousands of megapixels, through the utilization of DNN models in space is an essential solution. However, executing standard models directly on these satellite images may lead to excessive memory and potentially exhausting the available memory. This is particularly challenging in typical space environments where the memory capacity is insufficient for handling such large-scale images. Prior work \cite{denby2020} divided the image into several tiles maximizing inference accuracy at the expense of execution overhead. Additionally, downsampling to the input size of standard model architectures is frequently insufficient for achieving optimal performance \cite{van2018you}.

%With full knowledge of captured images, it determines the optimal tile size for each large-scale image and scales each tile to match the neural network's input size.

\begin{figure}[t]
\centering
\subfigure[xView \cite{lam2018xview}]{
\begin{minipage}[t]{0.44\linewidth}
\centering
\includegraphics[width=\linewidth]{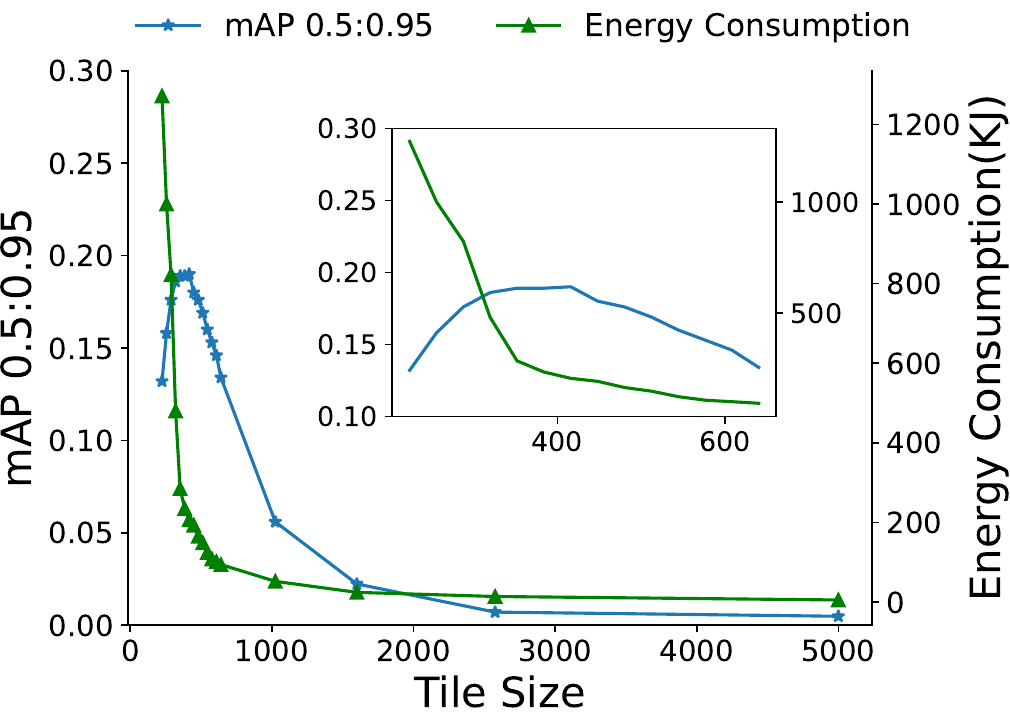}
%\caption{xView}
\end{minipage}%
}%
\subfigure[UAVOD10 \cite{han2022context}]{
\begin{minipage}[t]{0.44\linewidth}
\centering
\includegraphics[width=\linewidth]{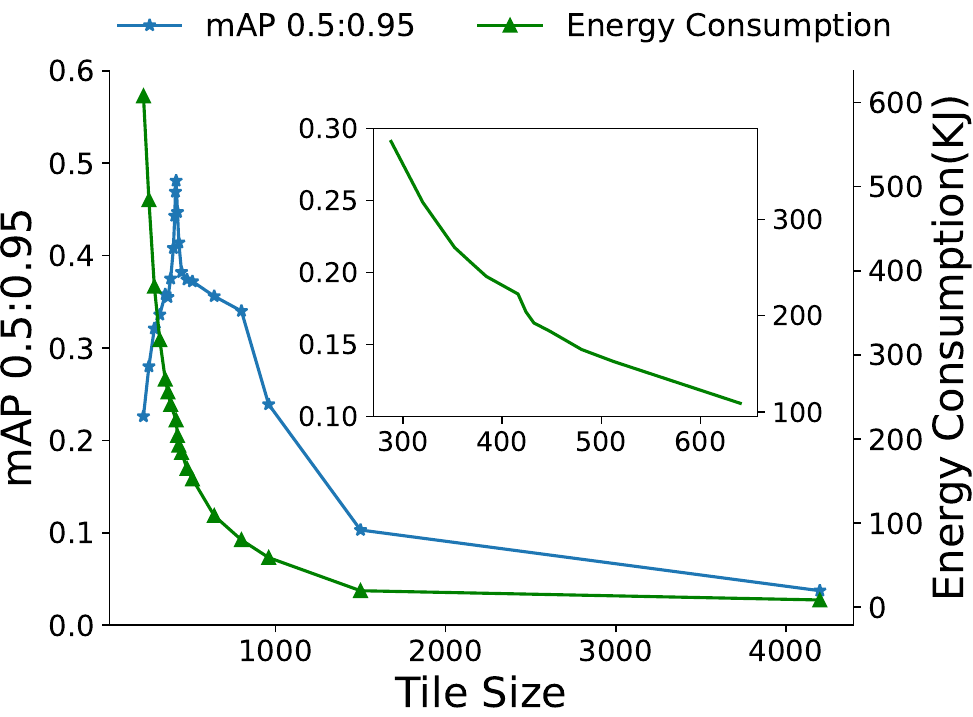}
%\caption{UAVOD10}
\end{minipage}%
}%
\centering
\caption{Effect of varied tile size on mAP accuracy and execution time of yoloV3-tiny model. mAP 0.5:0.95 is the average mAP over IoU threshold \cite{lin2014microsoft}.}
\label{fig:zz}
\end{figure}

A large image can be segmented into either a larger tile size with fewer tiles or a smaller tile size with numerous tiles.
After scaling each tile into input size of DNN counter, the execution time per tile remains constant.
Theoretically, opting for a larger tile size reduces image processing time, as each tile undergoes less degradation. Conversely, selecting a smaller tile size increases processing time, and each tile also undergoes less degradation.
To explore the impact of tile size on both inference accuracy and execution overhead, we conducted measurements on two datasets, as shown in Fig.~\ref{fig:zz}. Interestingly, both datasets display similar curves. As the tile size increases, the execution time decreases due to a smaller number of tiles per image. However, there is an optimal tile size that maximizes accuracy. Accuracy tends to deteriorate when the tile size deviates from this optimal size. This observation aligns with findings from previous studies \cite{denby2020,denby2023kodan}. 
The optimal tile size enables substantial improvements in accuracy while maintaining acceptable time. Moreover, the optimal tile size is not constant but varies depending on the DNN counters and image input size. Consequently, we determine the optimal tile size based on the combination of satellite image and DNN counter. Considering the trade-off between geospatial analysis accuracy (i.e., mAP accuracy) and execution overhead (i.e., processing time), we aim to identify an optimal tile size that aligns with the input size of the DNN counter.

We present a detailed approach for optimizing image size in tile-based processing using
Algorithm~\ref{alg:algorithm1}. 
The process initiates by initializing the tile sizes.
Drawing from the measurement results before NN counter deployment, we promptly narrow down the search interval and empirically establish the minimum and maximum tile sizes. We then iterate until the first image size meets the preset empirical size difference threshold.
%(i.e., $\epsilon$ is set to 1e-6).
Specifically, we divide the search interval into three equal fractions and compare the mAP accuracy at different sizes to identify optimal search intervals. The optimal tile size lies in the interval [$s_{midl}$, $s_{right}$] when $ mAP_{s_{left}} < mAP_{s_{right}}$, and vice versa. Finally, we obtain an approximate optimal tile size by taking the midpoint of the interval.
This method achieves a balance between accuracy and computational efficiency, providing a user-friendly solution with improved speed and accuracy of counting application.

\begin{algorithm}[t]
    \caption{Optimal Tile Size Selection}
    \label{alg:algorithm1}    
    \LinesNumbered
        \KwIn {minimum size $s_{left}$, maximum size $s_{right}$, threshold $\epsilon$, $mAP_{s_{left}}$, $mAP_{s_{right}}$} %%input
        \KwOut {optimal size $s_{best}$}    %%output
        
        $s_{left} \leftarrow s_{min}$;~~$s_{right} \leftarrow s_{max}$\;
        \While{$s_{right}$ $-$ $s_{left}$ \textgreater $\epsilon$}
            {$s_{midl} \leftarrow s_{left} + (s_{right} - s_{left})/3$\;
            $s_{midr} \leftarrow s_{right} - (s_{right} - s_{left})/3$\;
              \eIf {$ mAP_{s_{left}} \textless  mAP_{s_{right}}$}
                {$s_{left} \leftarrow s_{midl}$\;}
                {$s_{right} \leftarrow s_{midr}$\;}}
        $s_{best} = (s_{left} + s_{right})/2$\;
        Return $s_{best}$
\end{algorithm}

\subsection{Clustering-based Data Deduplication}
\begin{comment}
\begin{figure*}[htbp]
\centering
\subfigure[]{
\begin{minipage}[t]{0.2\linewidth}
\centering
\includegraphics[width=\linewidth]{}
\end{minipage}%
}
\subfigure[]{ 
\begin{minipage}[t]{0.2\linewidth}
\centering
\includegraphics[width=\linewidth]{}
\end{minipage}%
}%
\subfigure[]{
\begin{minipage}[t]{0.2\linewidth}
\centering
\includegraphics[width=\linewidth]{}
\end{minipage}
}%
\subfigure[]{
\begin{minipage}[t]{0.2\linewidth}
\centering
\includegraphics[width=\linewidth]{}
\end{minipage}
}%
\centering
\caption{}
\label{fig:mm}
\end{figure*}
\end{comment}

\begin{figure}[t]
\centering
\subfigure[]{
\begin{minipage}[t]{0.35\linewidth}
\centering
\includegraphics[width=\linewidth]{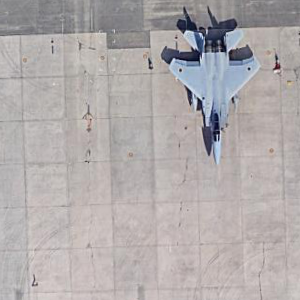}
\label{fig:1}
\end{minipage}%
}%
\subfigure[]{ 
\begin{minipage}[t]{0.35\linewidth}
\centering
\includegraphics[width=\linewidth]{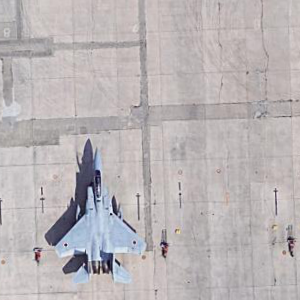}
\label{fig:2}
\end{minipage}%
}%
\centering
\caption{The tiles generated through image tiling in the DOTA \cite{9560031} dataset we utilize contain several images with similarities.}
\label{fig:mw1}
\end{figure}

\sys is tasked with classifying each tile into geographic contexts while complying with the computational constraints imposed by satellite hardware. 
A geographic context refers to a subset of images characterized by a high degree of similarity, along with geographic and transformation features. It is common for these images to exhibit a substantial degree of similarity, often remaining relatively static over time.
EO satellites periodically pass over identical locations on Earth's surface, capturing images that  exhibit significant similarity or near-identical characteristics at different times along their orbital path \cite{roy2014landsat}. As shown in Fig.~\ref{fig:mw1}, the two tiles acquired after tiling include an identical number of similar images. 
This is due to the short average revisit cycle of each satellite, such as GF-3, which revisits the same area at least twice every day, enabling the capture of the same geographical area multiple times \cite{gF}.

The presence of numerous semantically similar images exerts substantial pressure on the limited computational capacity within a satellite. This heightened demand for processing images may pose a computational challenge. To tackle this challenge, we propose a data deduplication strategy that involves processing representative tiles based on geographic context, rather than processing all similar tiles.
Certain images are computationally less demanding in specific contexts than in others. Due to the high predictability of satellite orbits, determining the contexts can be readily achieved. However, there may be image tiles for which the contexts are not immediately apparent, posing a challenge in their generation.

The technique efficiently generates contexts for image tiles by dividing them into multiple contexts. To cluster the representative image tiles based on similarity, the technique utilizes a low-dimensional label vector indicating the geographic features described by computing moments \cite{keen2005color} present in each image tile. The technique creates a set of contexts by performing \textit{k}-means clustering while exploring the Euclidean distance of the label vectors to measure similarity. Moreover, to enhance the DNN counter’s robustness to diverse sensors, we consider geographic label transformations like translations and rotations as objects may have arbitrary headings between 0 and 360 degrees. We also explore a range of cluster counts when partitioning the dataset into several clusters. Further investigation of this hyperparameter space represents an exciting avenue for future research.

\subsection{Bandwidth-aware downlinking throttling}
The downlinking bottleneck also poses a significant challenge in in-orbit counting, impacting the overall performance.
When receiving the tiles with confidence thresholds, \sys employs selection logic to determine the policy for handling these tiles. The optimal policy ensures that downlinking remains within the bandwidth budget constraint while transmitting as many tiles as possible to the ground to minimize counting errors.

The selection logic, based on the confidence threshold from the space-based DNN counter, is categorized into three groups (in Fig.~\ref{fig:123}): when the confidence threshold is relatively smaller (i.e., \textit{$<conf_{p}$}), \sys discards them directly; when confidence threshold is large enough (i.e., \textit{$>conf_{q}$}), \sys accepts the counting result; only when confidence threshold is between \textit{$conf_{p}$} and \textit{$conf_{q}$} (i.e., $[conf_{p},conf_{q}]$), \sys downlinks the tiles and processes them on the ground DNN counter. 
To comprehensively explore how the confidence threshold affects $CMAE$ (i.e., the mean difference between the estimated count and the ground truth), we vary the confidence threshold $conf_{p}$ under different contact times between satellite and ground (i.e., different downlinking data volume). As the tiles are sorted by the confidence threshold, and the objective is to downlink as many tiles as possible, we must consider the following three methods based on the choice of confidence thresholds when downlinking the tiles within $[conf_{p},conf_{q}]$ under the limited bandwidth budget constraint:

\begin{figure}[t]
\centering
\subfigure[Contact time is 6 mins.]{
\begin{minipage}[t]{0.5\linewidth}
\centering
\includegraphics[width=\linewidth]{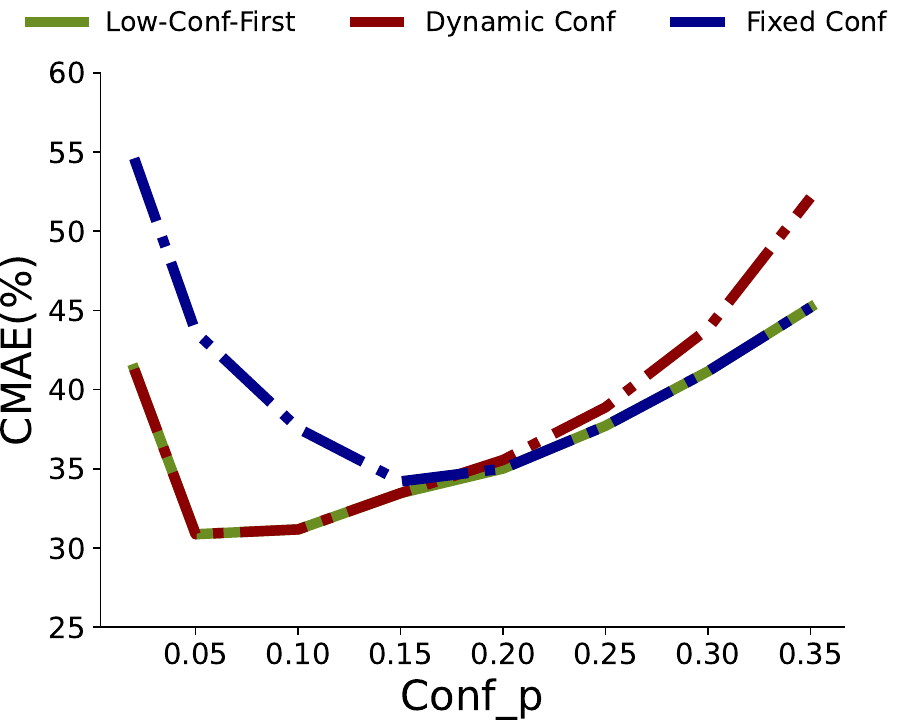}
\label{fig:1}
\end{minipage}%
}%
\subfigure[Contact time is 7 mins.]{ 
\begin{minipage}[t]{0.5\linewidth}
\centering
\includegraphics[width=\linewidth]{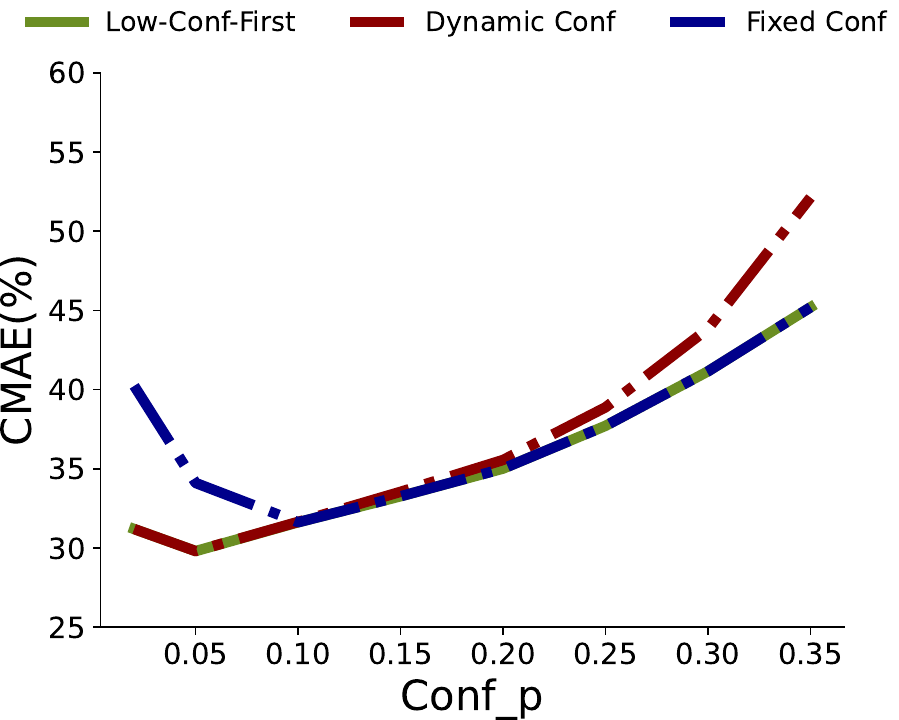}
\label{fig:2}
\end{minipage}%
}%

\subfigure[Contact time is 8 mins.]{
\begin{minipage}[t]{0.5\linewidth}
\centering
\includegraphics[width=\linewidth]{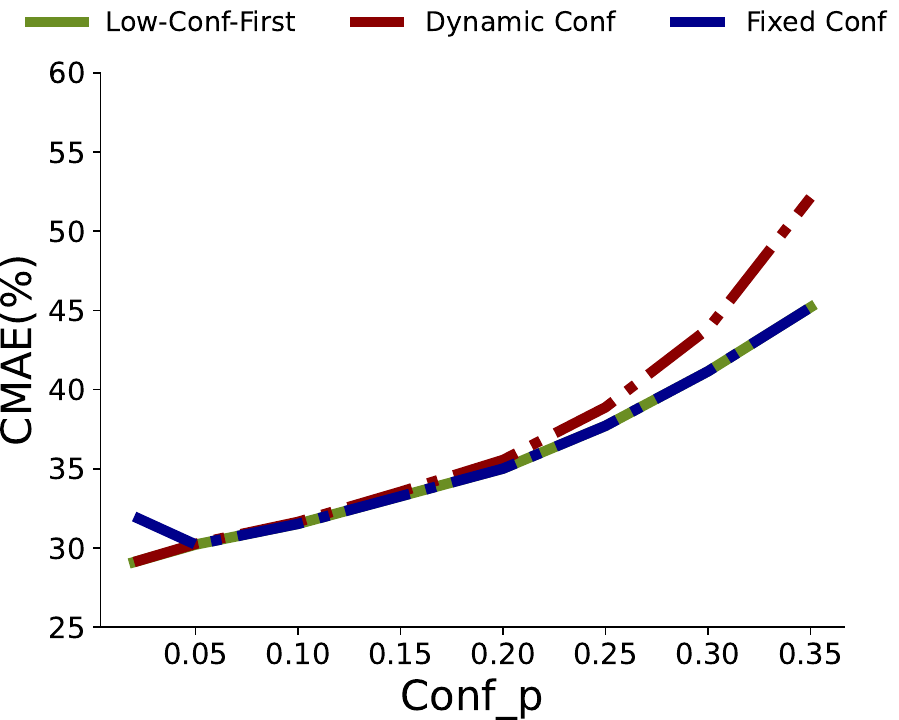}
%\caption{fig2}
\end{minipage}
}%
\subfigure[Contact time is 15 mins.]{
\begin{minipage}[t]{0.5\linewidth}
\centering
\includegraphics[width=\linewidth]{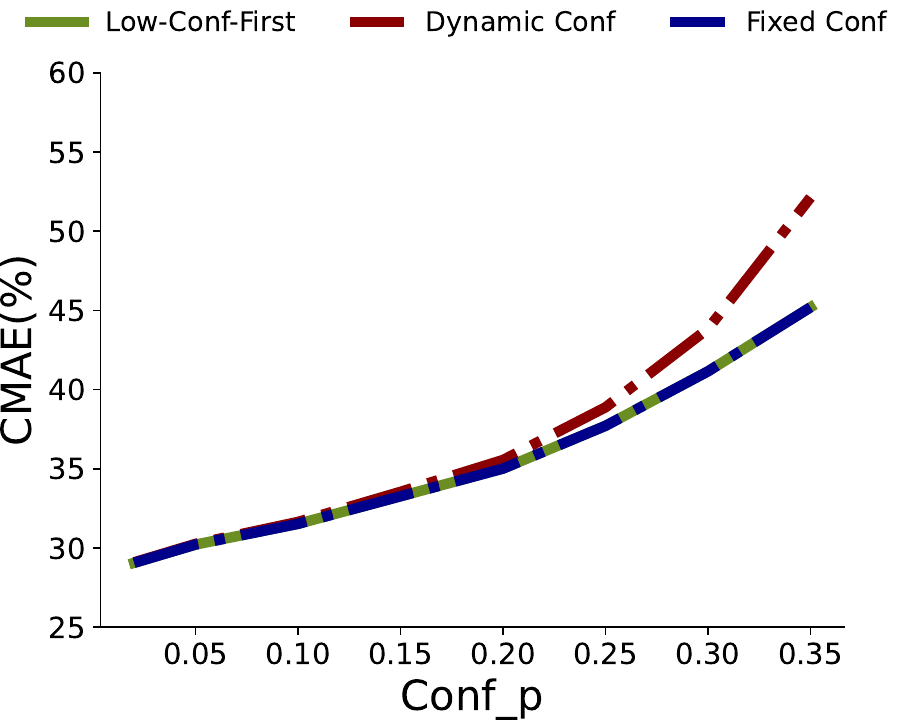}
%\caption{fig2}
\label{fig:mmw}
\end{minipage}
}%
\centering
\caption{A comparison of $CMAE$ (in Section \uppercase\expandafter{\romannumeral4}) in three methods based on the choice of confidence thresholds when downlinking in different contact times.}
\label{fig:mw}
\end{figure}

\begin{itemize}
    \item [$\bullet$] \textit{Low-Conf-First:} If there are still tiles remaining when the bandwidth is exhausted, we proceed to directly count these tiles and downlink the results.
    \item [$\bullet$] \textit{Fixed $Conf$:} If there are still tiles remaining when the bandwidth is exhausted, we only count tiles whose confidence thresholds are higher than the fixed $conf_{q}$.
    \item [$\bullet$] \textit{Dynamic $Conf$:} We first count the tiles with confidence thresholds higher than the preset $conf_{q}$ and then count the tiles whose confidence thresholds are within $[conf_{p},conf_{q}]$ until the bandwidth is exhausted. The value of $conf_{q}$ varies depending on the downlinking constraint.
\end{itemize}

The observations from Fig.~\ref{fig:mw} are summarized as follows: 

\begin{itemize}
\item [$\bullet$] \textbf{Dynamic $conf_{p}$ leads to performance improvement.} 
When the downlink capacity is insufficient, both \textit{Low-Conf-First} and \textit{Dynamic $Conf$} show similar performance, as all remaining tiles are counted and \textit{Dynamic $Conf$} is not effective.
In this scenario, a larger $conf_{p}$ leads to a lower $CMAE$, as it facilitates the downlinking of more high-confidence tiles.
Moreover, when downlink capacity is sufficient, both \textit{Fixed $Conf$} and \textit{Dynamic $Conf$} exhibit comparable performance, as all tiles are counted. Consequently, \textit{Fixed $Conf$} is not enabled.
In this case, a larger $conf_{p}$ results in higher $CMAE$, as it discards the high-value tiles with confidence thresholds below $conf_{p}$.
Therefore, choosing an appropriate dynamic $conf_{p}$ is crucial for improving counting performance. It motivates to downlink high-confidence images first and downlink some low-confidence images within the available bandwidth.

\item [$\bullet$] \textbf{Optimal $conf_{p}$ improves performance.} 
When downlink capability is sufficient and $conf_{p}$ increases to a certain value, all methods discard the tiles below $conf_{p}$, resulting in a counting error.
Here both \textit{Low-Conf-First} and \textit{Fixed $Conf$} exhibit identical performance, since \textit{Fixed $Conf$} fails to work effectively. Additionally, \textit{Dynamic $Conf$} incurs a higher counting error, as counting tiles larger than $conf_{q}$ in space is not as accurate as on the ground. Therefore, optimizing $conf_{p}$ is crucial.

\item [$\bullet$] \textbf{$conf_{q}$ alleviates the downlink constraint without greatly affecting performance.} 
In Fig.~\ref{fig:mmw}, when the downlink capability is sufficient (i.e., dynamic $conf_{q}$ does not change), and the initial $conf_{q}$ is not too large (i.e., smaller than 0.2), \textit{Dynamic $Conf$} and \textit{Low-Conf-First} show similar performance. However, \textit{Dynamic $Conf$} has the advantage of increasing the downlink volume compared to \textit{Low-Conf-First}.
\end{itemize}

Therefore, strategically selecting the optimal confidence threshold is crucial to downlink more high confidence tiles. However, we face the challenge of determining the downlinking method for confidence thresholds and fully exploiting the bandwidth budget.
Algorithm \ref{alg:algorithm2} describes the bandwidth-aware downlinking throttling procedure, which takes the bandwidth requirement and tiles obtained by tiling and clustering (as shown in Fig.~3) as input and produces a set of tiles that can be downlinked by the scarce bandwidth constraint.
We first identify all the clustered tile sets on the satellite and discard tiles with a confidence threshold lower than a specified empirically $conf_{p}$ (lines 5-6). Tiles with high confidence $conf_{q}$ are included directly in the $C_{space}$ (lines 7-8). We calculate the current remainder bandwidth and maximize it to downlink images (lines 13-18), In other words, the remaining tiles are sorted based on data size in descending order, and the count results are added to the transmitted tile set $S_{trans}$ if sufficient available bandwidth is present.

\begin{algorithm}[!t]
    \caption{Bandwidth-aware downlinking throttling}
    \label{alg:algorithm2}
    \LinesNumbered
        \KwIn {minimum confidence $conf_{p}$, maximum confidence $conf_{q}$, bandwidth requirement $Band_{max}$, filtered ROI set $tiles$} %%input
        \KwOut {transmitted tile set $S_{trans}$, count from space $C_{space}$}    %%output
        
        %\STATE  \yx{initialize $conf_{tiles}$}\; 
        Initialize $S_{trans}$, $C_{space}$, $conf_{tiles}$\;
        $Band_{rest} \leftarrow Band_{max}$;\tcp{Remaining bandwidth assignment}
        % \WHILE{$conf_{tiles}$  $conf_{q}$}
        %     \STATE $S_{tiles}.add()$\;
        % \ENDWHILE
        
        \ForEach{$tile$ \textbf{in} $tiles$}
        {$scores,labels \leftarrow Geospatial \ DNN \ Counter(tile)$\;
        \If{$scores.mean()$ $\textless$ $conf_{p}$}
            {$continue$; }
        \eIf{$scores.mean()$ $\textgreater$ $conf_{q}$} 
            {$C_{space} += C_{tile}$;}
            {$conf_{tiles}.add()$;} }
        {$conf_{tiles}.sort()$;\tcp{Sort by confidence score}}
       \ForEach{$tile$ \textbf{in} $conf_{tiles}$}
        {$Size \leftarrow  tile.size()$;\tcp{Measure size of tile}
        \eIf{$Band_{rest}.size()$ $\geq$ $Size$}
            {$S_{trans}.add()$;}{$break$;}}
        Return $S_{trans}$, $C_{space}.$
\end{algorithm}

\section{evaluation}

\subsection{Methodology} 

%%%%第一段完成了
\subsubsection{Heterogeneous computational hardware}~A computational satellite enhances onboard sensing, communications, and control operations.
In this context, the tested satellites are equipped with two industrial modules: Raspberry Pi 4B (RPI4) and Atlas 200 DK (Atlas). These modules are preferred for their cost-effectiveness and programming simplicity. This work focuses on characterizing onboard computing utilizing these two modules, recognized as widely-used computational hardware in satellite applications. 

%介绍功率

\subsubsection{ROI-based instance selection}~To address redundant counting in small tiles containing invalid information, such as backgrounda, recognized challenge in spatial vision, we utilize established techniques for regions of interest (ROI) \cite{ROI}, as depicted in Fig.~\ref{fig:123}. We also implement non-maximal suppression to the global matrix of bounding box predictions to alleviate overlapping detections \cite{bailo2018efficient}. 
The operation is compatible with this work, processing in-orbit images to save bandwidth and aligning seamlessly with our core contributions in producing statistical counting results.

\begin{table}[]
\centering
\begin{tabular}{llll}
\hline
%\cer
Datasets & Size & \begin{tabular}[c]{@{}l@{}}GSD (m)\end{tabular} & \begin{tabular}[c]{@{}l@{}}Volume (GB)\end{tabular} \\ \hline
xView   & 3000           & 0.3           & 20   \\
DOTA    & 4000           & 0.1$\sim$0.81 & 34.3 \\
UAVOD10 & 1000$\sim$4800 & 0.15          & 0.9  \\ \hline
\end{tabular}
\caption{Datasets in the evaluation. Size: large-scale geospatial image resolution. GSD: geographic distance between adjacent pixels.}
\label{lab:pp}
\end{table}

\subsubsection{Experiment Setups}~Theoretical satellite-to-ground bandwidth can achieve 100 Mbps, matching the air interface transmission rate \cite{oo}. However, real-world measured satellite-ground bandwidth is limited to 30-50 Mbps due to reception losses. Additionally, the default satellite-ground contact time is set at 6 minutes, and the default DNN counters on the satellite and ground are YOLOV3-tiny and YOLOV3, respectively.

\begin{comment}
\begin{figure}[t]
   \centering
   \setlength{\belowcaptionskip}{-0.5cm}
   \begin{center}
     \includegraphics*[width=0.6\linewidth]{graphics/paper_testbed.pdf}
     %\vspace{-1.5cm}
   \caption{Energy-consumpting hardware device in the tested satellite payloads.}
   \label{fig:43}
   \end{center}
\end{figure}
\end{comment}

\subsubsection{Datasets}~We utilize three geospatial datasets covering various scenes, including buildings, planes, and tracks (in Table \ref{lab:pp}), for evaluation. Each satellite captures images along its ground track, and although object counts may exhibit high temporal correlation, there are currently no publicly available datasets covering multiple days and providing diverse object contexts with sufficient instances. To address this limitation, we employed rotations and data augmentation. We simulated image captures as the satellite passed over its ground track by flipping and rotating 50\% of the images in the dataset.

\subsubsection{DNN counters and ground truth counts}~In Table~\ref{lab:PDC}, we present publicly available DNN counters customized for each data sample, primarily small or medium-sized to accommodate the resource-constrained nature of satellites. The evaluation is conducted based on the ground truth returned by the dataset.

\subsubsection{Metrics}~To quantify the performance of the counting system, we use \textit{Count Mean Absolute Error (CMAE)}, defined as $ \sum \left|y_{i}-g_{i}\right|/\sum g_{i}$, where $y_{i}$ and $g_{i}$ are our counter count and true count, respectively. The metric indicates the extent to which our approximate counts deviate from the ground truth, and a narrower $CMAE$ is considered better. In addition, we leverage the data size to demonstrate changes in the volume of data, offering an indirect reflection of bandwidth utilization.

%In addition, we validate our system using generic detectors to ensure ease of experiments and result in reproducibility.

\begin{table}[t]
\centering
\begin{tabular}{@{}lll@{}}
\hline
%\cer
DNN Counters     & Input & mAP \\ \hline
\yolo (Ground)&      416*416  &   55.3  \\
\yolo-tiny     &    416*416   &   33.1  \\
%yolofastestv2    &   352*352    &   24.1  \\
ssd mobilenetv2    &  200*300     &   22.0  \\ \hline
\end{tabular}
\caption{The DNN counters architecture used in this work. mAP: mAP accuracy on COCO dataset \cite{lin2014microsoft}.}
\label{lab:PDC}
\end{table}

\subsubsection{Baselines}~We compare \sys to four baseline methods:
(1) \textit{Space-Only:} All images are processed using the onboard DNN counter, and the resulting counting data is transmitted to the ground;
(2) \textit{Ground-Only \cite{denby2020}:} Satellites operate as “bent-pipe”, collecting images and downlinking them to the ground; Due to bandwidth limitations, the approach is to downlink as many observations as possible within the limited contact time;
(3) \textit{TIANSUAN \cite{wang2023tiansuan}:} Following onboard counting, satellites exclusively transmit results with confidence thresholds surpassing a predetermined empirical threshold, downlinking the remaining data to the ground within the constrained contact time;
(4) \textit{Kodan \cite{denby2023kodan}:} The images processed in orbit are categorized into different levels, with a preference for transmitting high-value pictures. Note that while we share similarities with \textit{Kodan} in onboard processing, there is a distinction in transmitting — \textit{Kodan} does not consider bandwidth limitations.
Theoretically, the outcome obtained from \textit{Kodan} functions as an upper bound.
%\new{Figure 8, 11, 13, 15 has added the kodan baseline, but only when bandwidth is insufficient, we surpass kodan. In the ideal case that bandwidth is sufficient, our method is not as good as it. Would you please comment on whether to add kodan comparison, and also compare with kodan in related work?}

\subsection{End-to-End performance}

% \begin{figure}[t]
% \centering
% \subfigure[xView.]{
% \begin{minipage}[t]{0.5\linewidth}
% \centering
% \includegraphics[width=\linewidth]{graphics/fig10_xView_bandwidth}
% %\caption{fig1}
% \end{minipage}%
% }%
% \subfigure[UAVOD10.]{
% \begin{minipage}[t]{0.5\linewidth}
% \centering
% \includegraphics[width=\linewidth]{graphics/fig10_UAVOD10_bandwidth}
% %\caption{fig2}
% \end{minipage}%
% }%
% \centering
% \caption{Effect of varied downlink bandwidth on $CMAE$. Note that the contact time is set to 6 minutes.}
% \label{fig:mea1_2}
% \end{figure}

% \begin{figure}[t]
% \centering
% \subfigure[xView.]{
% \begin{minipage}[t]{0.5\linewidth}
% \centering
% \includegraphics[width=\linewidth]{graphics/fig10_xView-konda}
% %\caption{fig1}
% \end{minipage}%
% }%
% \subfigure[UAVOD10.]{
% \begin{minipage}[t]{0.5\linewidth}
% \centering
% \includegraphics[width=\linewidth]{graphics/fig10_UAVOD10-konda}
% %\caption{fig2}
% \end{minipage}%
% }%
% \centering
% \caption{Effect of varied downlink bandwidth on $CMAE$. Note that the contact time is set to 6 minutes.}
% \label{fig:mea1_22}
% \end{figure}

\subsubsection{\sys enhances performance}
\begin{figure}[t]
   \centering
   \setlength{\belowcaptionskip}{-0.5cm}
   \begin{center}
     \includegraphics*[width=0.9\linewidth]{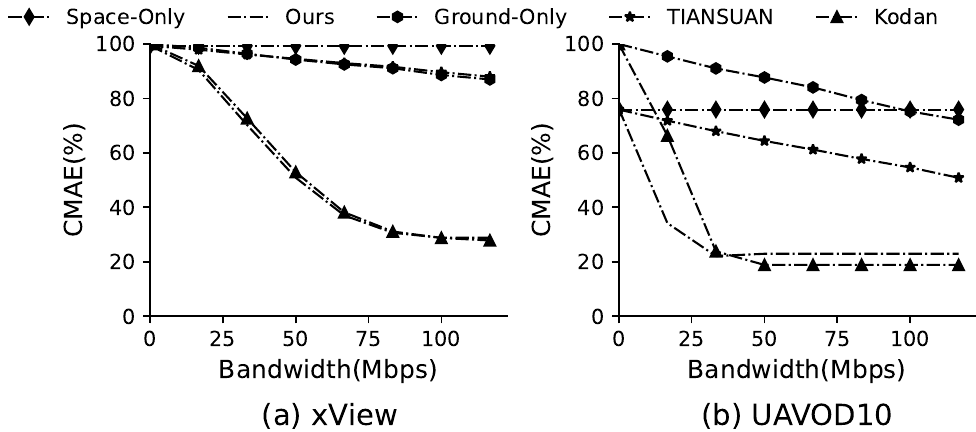}
   \caption{The performance of $CMAE$ on the varying satellite-ground bandwidth.}
   \label{fig:mea1_22}
   \end{center}
\end{figure}

Fig.~\ref{fig:mea1_22} illustrates the experimental results with the varying bandwidths.
In all methods, except for \textit{Space-Only}, $CMAE$ decreases as more bandwidth resources are utilized and more images are downlinked. 
\sys outperforms the vanilla baseline TIANSUAN, which relies on a fixed confidence threshold, by reducing the counting error by 1.93$\times$ and 2.51$\times$ on average across different datasets. 
This improvement is achieved because \sys intelligently selects representative and high-value images for downlinking, leading to enhanced counting performance. Additionally, \sys achieves a 9.6$\times$ bandwidth-efficient improvement compared to TIANSUAN under limited bandwidth constraints.

Counting performance is subject to the characteristics and composition of the dataset used for evaluation. Despite variations in the datasets, \sys shows similar performance to \textit{Kodan}, with differences mainly attributable to the downlinking. The key advantage of our system over \textit{Kodan} lies in its incorporation of bandwidth-aware downlinking throttling. 
In scenarios with extremely scarce bandwidth, especially close to real-world values (i.e., below 50 Mbps in Fig.~\ref{fig:mea1_22}(b)), 
\sys enables efficient bandwidth allocation and directing the remaining resources to other applications.

\subsubsection{\sys provides effective counting with limited energy constraints}
\begin{figure}[t]
   \centering
   \setlength{\belowcaptionskip}{-0.5cm}
   \begin{center}
     \includegraphics*[width=0.5\linewidth]{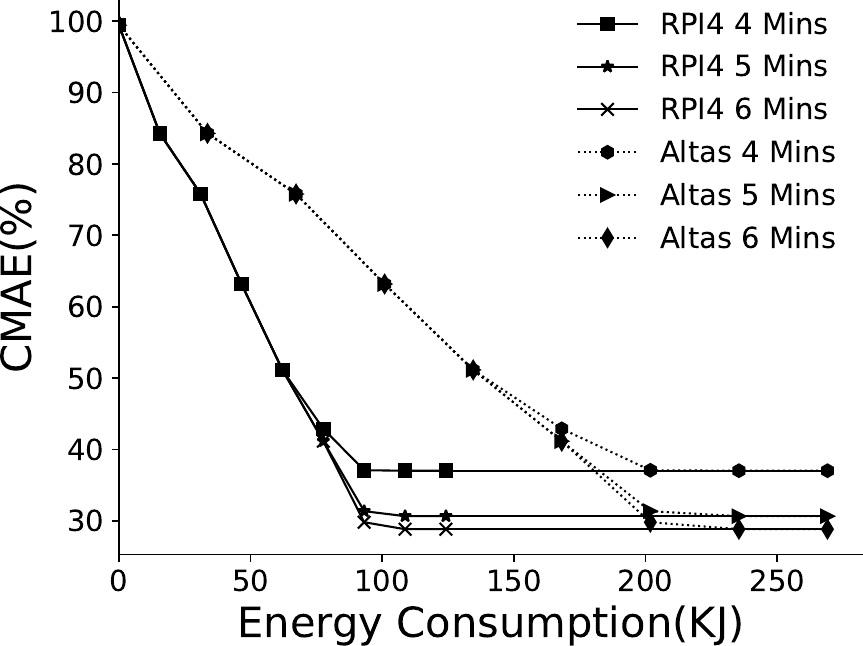}
   \caption{\sys operates on COTS hardware with varying computational energy budgets, which is a fraction of the total solar energy available.}
   \label{fig:energy}
   \end{center}
\end{figure}

Fig.~\ref{fig:energy} explores the impact of computational energy on the performance of \sys, considering various hardware configurations and contact times. 
Each satellite is theoretically restricted to a daily energy collection of up to 260KJ, allocated to computing operations within a specified energy budget. Results show that, under identical computational energy constraints, longer contact time leads to lower $CMAE$ as the increased downlinking of images to the ground. Additionally, both hardware setups achieve comparable CMAE values within the same contact time. However, RPI4 (with 6W power) featuring computing-limited hardware outperforms Atlas (with 13W power) by saving approximately 50\% of energy. This advantage stems from the RPI4's ability to efficiently process more data, thus minimizing the $CMAE$.
%Given the energy-intensive vision tasks in space, such energy savings with low computing power could significantly extend the battery life.

\subsubsection{Low-power hardware also improves performance}
\begin{figure}[t]
   \centering
   \setlength{\belowcaptionskip}{-0.5cm}
   \begin{center}
     \includegraphics*[width=0.9\linewidth]{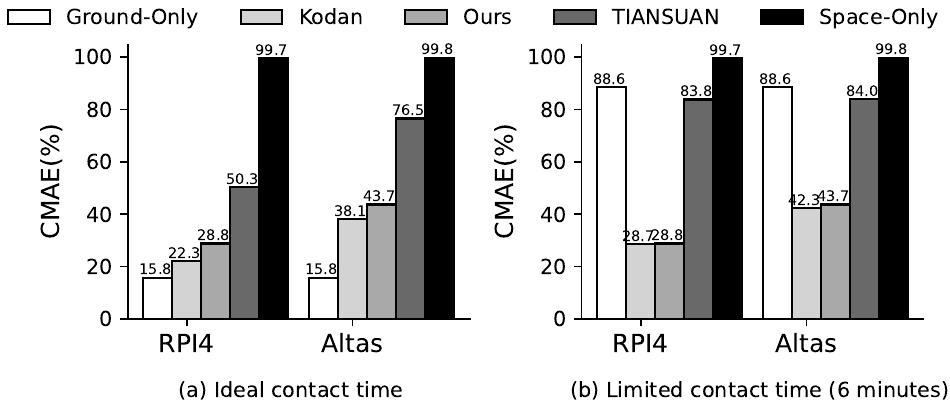}
   \caption{A comparison of $CMAE$ on different hardware with different contact times (dataset: xView, energy budget: 150KJ/day, onboard counter: \yolo-tiny).}
   \label{fig:hard22}
   \end{center}
\end{figure}

The performance of \sys is relevant with accelerators running DNN counters, as shown in Fig.~\ref{fig:hard22}. Compared to Altas, RPI4 significantly reduces $CMAE$ by 34\%, primarily attributed to its ability to process and downlink more images per track. 
Specifically, for the same $CMAE$, RPI4 requires a shorter contact time, resulting in less downlinked data; with the same contact time, RPI4 can achieve a lower $CMAE$. 
% \begin{figure}[t]
%    \centering
%    \subfigure[Ideal contact time.]{
%     \begin{minipage}[t]{0.5\linewidth}
%     \centering
%     \includegraphics[width=\linewidth]{graphics/hardware-ideal.pdf}
%     \end{minipage}%
%     }%
%    \subfigure[Limited contact time (6 minutes).]{
%     \begin{minipage}[t]{0.5\linewidth}
%     \centering
%     \includegraphics[width=\linewidth]{graphics/hardware-real.pdf}
%     \end{minipage}%
%     }%
%    \caption{\sys on different hardware with different contact times (dataset: xView, energy budget: 150KJ/day, onboard counter: \yolo-tiny).}
%    \label{fig:hard2}
%    %\vspace{-0.8cm} 
% \end{figure}

% \begin{figure}[t]
%    \centering
%    \subfigure[Ideal contact time.]{
%     \begin{minipage}[t]{0.5\linewidth}
%     \centering
%     \includegraphics[width=\linewidth]{graphics/hardware-ideal-konda.pdf}
%     \end{minipage}%
%     }%
%    \subfigure[Limited contact time (6 minutes).]{
%     \begin{minipage}[t]{0.5\linewidth}
%     \centering
%     \includegraphics[width=\linewidth]{graphics/hardware-real-konda.pdf}
%     \end{minipage}%
%     }%
%    \caption{\sys on different hardware with different contact times (dataset: xView, energy budget: 150KJ/day, onboard counter: \yolo-tiny).}
%    \label{fig:hard22}
%    %\vspace{-0.8cm} 
% \end{figure}

\begin{figure}[t]
   \centering
   \setlength{\belowcaptionskip}{-0.5cm}
   \begin{center}
     \includegraphics*[width=0.75\linewidth]{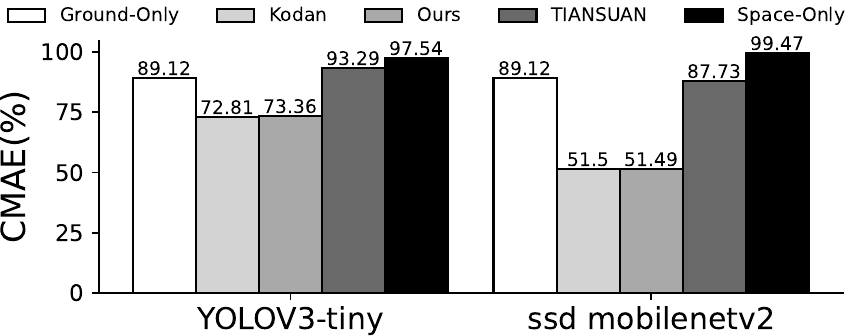}
   \caption{A comparison of counting performance on various onboard DNN counters under the limited bandwidth (50 Mbps).}
   \label{fig:DNN1}
   \end{center}
\end{figure}
\subsubsection{Exploiting diverse DNN counters}
Fig.~\ref{fig:DNN1} illustrates the counting performance of different DNN counters under the same energy constraint. Despite the diverse choices of counters on the satellite, their performance remains similar. Both \sys and $Kodan$ exhibit very comparable performance, as both methods select less accurate DNN counters in space, with the ground counter providing more accurate counts. This observation highlights the significance of adaptive image tiling, as it enables different selection logics for each tile based on confidence thresholds. Execution with a smaller tile size (e.g., ssd mobilenetv2) experiences less degradation, resulting in a narrower $CMAE$.

\subsubsection{Exploiting diverse datasets}
\begin{figure}[t]
   \centering
   \setlength{\belowcaptionskip}{-0.5cm}
   \begin{center}
     \includegraphics*[width=0.75\linewidth]{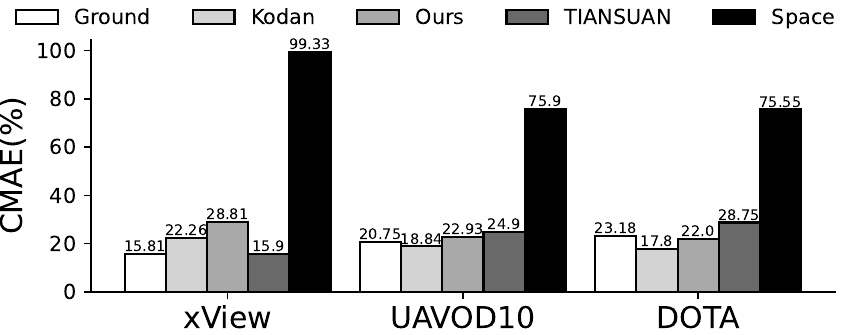}
   \caption{A comparison of counting performance on three datasets under the unlimited downlinking.}
   \label{fig:mae1}
   \end{center}
\end{figure}

Fig.~\ref{fig:mae1} shows the counting performance on widely-used datasets, comparing the baselines with \sys under unlimited downlinking to the ground. 
The performance of \textit{TIANSUAN} consistently varies across different datasets, as this method depends on fixed empirical confidence thresholds, thereby affecting counting performance.
\sys reduces counting error by 3.4$\times$ on average, compared to \textit{Space-Only}. 
With \textit{Ground-Only}, computational constraints are further mitigated and ground counters can process all tiles with higher precision, resulting in the theoretically lowest  achievable $CMAE$.
\textit{Kodan} and \sys implement ROI-based instance selection, image deduplication, and onboard computing. However, these methods lead to a comparable performance with \textit{Ground-Only}.

%\sys overcomes downlink limitations by bandwidth-aware downlinking throttling, resulting in a comparable performance with SOTA methods.

\begin{comment}
\begin{figure}[t]
\centering
\subfigure[xView.]{
\begin{minipage}[t]{0.5\linewidth}
\centering
\includegraphics[width=\linewidth]{graphics/fig10_xView}
%\caption{fig1}
\end{minipage}%
}%
\subfigure[UAVOD10.]{
\begin{minipage}[t]{0.5\linewidth}
\centering
\includegraphics[width=\linewidth]{graphics/fig10_UAVOD10}
%\caption{fig2}
\end{minipage}%
}%
\centering
\caption{Effect of varied contact time on $CMAE$. Note that the downlink bandwidth is set to 100 Mbps.}
\label{fig:mea1}
\end{figure}
\end{comment}

% \begin{figure}[t]
%    \centering
%    \setlength{\belowcaptionskip}{-0.5cm}
%    \begin{center}
%      \includegraphics*[width=0.75\linewidth]{fig6}
%    \caption{A comparison of $CMAE$ ruDNNing \sys was performed on three datasets under the ideal scenario (100\% downlink to ground).}
%    \label{fig:mae}
%    \end{center}
% \end{figure}

% \begin{figure}[t]
%    \centering
%    \setlength{\belowcaptionskip}{-0.5cm}
%    \begin{center}
%      \includegraphics*[width=0.75\linewidth]{graphics/models.pdf}
%    \caption{\sys operates on various onboard DNN counters under the ideal scenario.}
%    \label{fig:DNN}
%    \end{center}
% \end{figure}

\subsection{Validation of Key Designs}
\subsubsection{Clustering-based data deduplication enhances bandwidth efficiency} In Fig.~\ref{fig:111}, performing \textit{Clustering} leads to a reduction in downlink data. Compared to \textit{No-Clustering}, the downlink volume in the \textit{Clustering} scenario is approximately 5.6\% less, accounting for 32.8\% of the volume observed in \textit{No-Clustering}. This is primarily due to the fact that, \textit{Clustering} transmits only representative similar or duplicated tiles to the ground after the clustering process. Additionally, within the downlinking constraints, \textit{Clustering} enables the transmission of additional small tiles to the ground, which can be integrated into the final counting result.

\subsubsection{Bandwidth-aware downlinking throttling improves performance}
\begin{figure}[t]
   \centering
   \subfigure[Data Deduplication.]{
    \begin{minipage}[t]{0.44\linewidth}
    \centering
    \includegraphics[width=\linewidth]{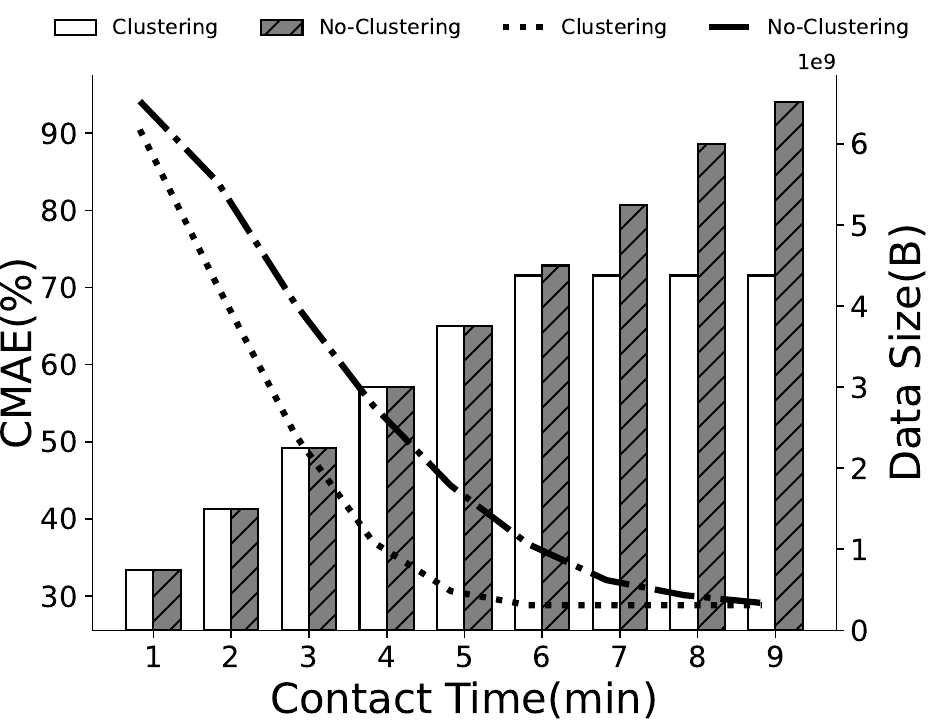}
    \label{fig:111}
    \end{minipage}%
    }%
   \subfigure[Downlinking throttling.]{
    \begin{minipage}[t]{0.44\linewidth}
    \centering
    \includegraphics[width=\linewidth]{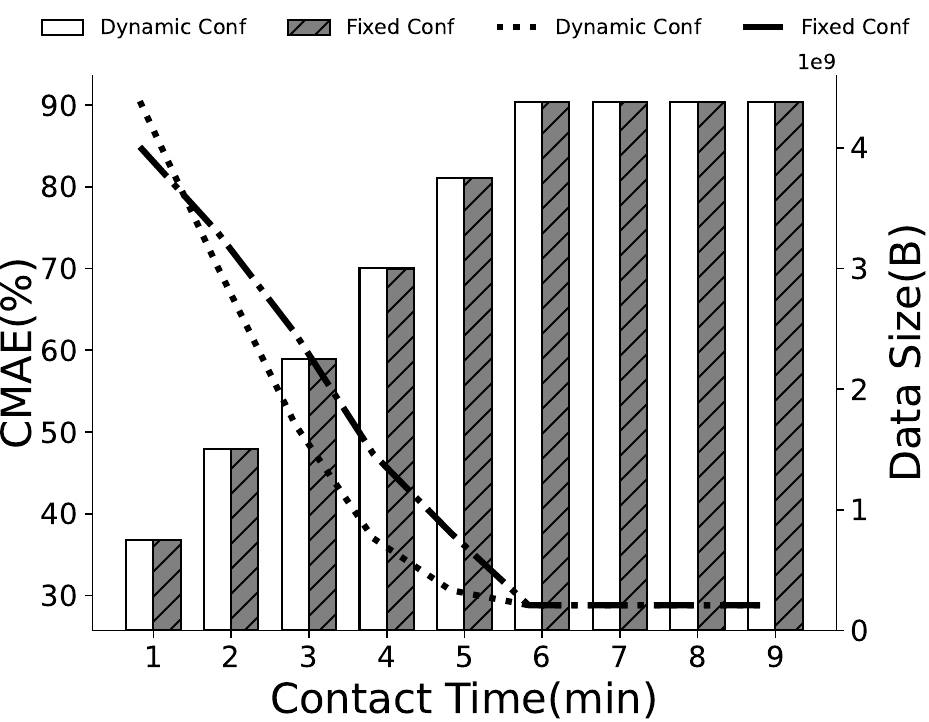}
    \label{fig:222}
    \end{minipage}%
    }%
   \caption{Ablation study on key designs.}
   \label{fig:33}
\end{figure}

Fig.~\ref{fig:222} demonstrates that \textit{Dynamic Conf} outperforms \textit{Fixed Conf} as the contact time increases.
When contact time is limited, \textit{Dynamic Conf} can selectively downlink high-confidence tiles, reducing the counting error compared to the indiscriminate downlinking of tiles in \textit{Fixed Conf}. 
Both \textit{Dynamic Conf} and \textit{Fixed Conf} exhibit similar performance when all the tiles can be downlinked.
\begin{comment}
Our downlink data is 0.3\%-1.6\% less than that of \textit{fixed threshold} designs, accounting for  2.4\%-20.4\% of the downlink data of the \textit{fixed confidence} approach.    
\end{comment}

\section{Related Work}

Satellite networking has witnessed substantial growth, with a predominant focus in research on inter-satellite networking \cite{handley2018delay,klenze2018networking}.
However, the challenges posed by the scarce downlink bandwidth and the associated bottleneck are also crucial.
OEC \cite{denby2020} focuses on the downlink bottleneck and shifts it to the inelastic computation capacities.
Another recent work, Kodan \cite{denby2023kodan} has proposed filtering low-value data and prioritizing high-value data for downlinking to mitigate the downlink bottleneck. However, Kodan considers the constraints of scarce satellite-ground bandwidth, a limitation that we aim to address in this work.
This work not only tackles the downlink bottleneck through bandwidth-aware downlinking throttling but also addresses the computational bottlenecks.

The computational bottleneck represents a major challenge for satellite systems. Some works, such as \cite{giuffrida2020cloudscout} and \cite{giuffrida2021varphi}, have explored the viability of utilizing DNN models for in-orbit processing. However, these approaches do not directly tackle the specific challenges addressed by \sys, such as operating with real-world energy budgets. Moreover, several works focus on optimizing DNN models for accuracy or speed in terrestrial applications \cite{xu2018deepcache,zhang2022comprehensive, xu2021video,cao2021thia,zhang2023comprehensive,zhou2023mixed,zhou2023video}.
Various terrestrial and embedded systems that operate on harvested energy \cite{xu2020approximate,nardello2019camaroptera} can transmit data at any time within energy constraints. However, this continuous data transmission capability is impractical for satellites as they can only transmit data when they are in proximity to ground stations. Moreover, the limited bandwidth available for satellite communication is significantly smaller than that of ground-based connections.

Vision tasks in EO satellites have been extensively studied and proven valuable for scientific investigations \cite{jiang2019crowd,chen2017rethinking,chattopadhyay2017counting}. These applications span various domains, including computer systems, satellite systems, satellite networks, and machine learning systems.
Developing a comprehensive scheduling system that considers image size, processing speed, energy constraints, and orbital mechanics poses a challenging research problem in computer systems. We are actively working to resolve the computer systems design challenges associated with satellite computing under real-world satellite constraints. 
Our system leverages a tradeoff between accuracy and execution time, and effectively addresses downlink bottleneck by bandwidth-aware downlinking throttling.
\section{Conclusion}

This work introduces an analytics system tailored for addressing object counting queries on EO satellites. \sys utilizes both less accurate in space and more accurate ground-based DNN models to determine earth object counts within the constraints of computation and communication. \sys is designed to minimize counting errors under energy and bandwidth constraints. Extensive experiments show that \sys can reduce counting error by 3.4$\times$ on average, compared to onboard computing.

%\section*{Acknowledgment}

\footnotesize
\bibliographystyle{IEEEtran}
\bibliography{ref}

\end{document}